\documentclass[12pt]{article}
\usepackage{amssymb,amsmath,graphicx,epsfig}
\def\issue(#1,#2,#3){{\bf #1}, #2 (#3)} 

\def\APP(#1,#2,#3){{\rm Acta Phys.\ Polon.} \ \issue({\bf #1},#2,#3)}
\def\ANP(#1,#2,#3){{\rm Annals of Physics} \ \issue({\bf #1},#2,#3)}
\def\ARNPS(#1,#2,#3){{\rm Ann.\ Rev.\ Nucl.\ Part.\ Sci.} \ \issue({\bf #1},#2,#3)}
\def\CPC(#1,#2,#3){{\rm Comp.\ Phys.\ Comm.} \ \issue({\bf #1},#2,#3)}
\def\CIP(#1,#2,#3){{\rm Comput.\ Phys.} \ \issue({\bf #1},#2,#3)}
\def\EPJ(#1,#2,#3){{\rm Eur.\ Phys.\ J.} \ \issue({\bf #1},#2,#3)}
\def\EPJD(#1,#2,#3){Eur.\ Phys.\ J. Direct\ C \ \issue({\bf #1},#2,#3)}
\def\IJMP(#1,#2,#3){{\rm Int.\ J.\ Mod.\ Phys.} \ \issue({\bf #1},#2,#3)}
\def\JHEP(#1,#2,#3){{\rm J.\ High Energy Physics} \ \issue({\bf #1},#2,#3)}
\def\JP(#1,#2,#3){{ J.\ Phys.} \ \issue({\bf #1},#2,#3)}
\def\MPL(#1,#2,#3){{Mod.\ Phys.\ Lett.} \ \issue({\bf #1},#2,#3)}
\def\NP(#1,#2,#3){{Nucl.\ Phys.} \ \issue({\bf #1},#2,#3)}
\def\NIM(#1,#2,#3){{ Nucl.\ Instrum.\ Meth.} \ \issue({\bf #1},#2,#3)}
\def\PL(#1,#2,#3){{ Phys.\ Lett.} \ \issue({\bf #1},#2,#3)}
\def\PR(#1,#2,#3){{ Phys.\ Rev.} \ \issue({\bf #1},#2,#3)}
\def\PRL(#1,#2,#3){{ Phys.\ Rev.\ Lett.} \ \issue({\bf #1},#2,#3)}
\def\SJNP(#1,#2,#3){{ Sov.\ J. Nucl.\ Phys.} \ \issue({\bf #1},#2,#3)}
\def\ZP(#1,#2,#3){{Zeit.\ Phys.} \ \issue({\bf #1},#2,#3)}
\def\be {\begin{equation}}
\def\ee {\end{equation}}
\def\bea {\begin{eqnarray}}
\def\eea {\end{eqnarray}}

\def\bbbar {B^0-\overline{B}{}^0}
\def\bsbsbar {B_s^0-\overline{B_s}{}^0}

\def\bra {\langle}
\def\ket {\rangle}
\begin{document}
\begin{titlepage}
\begin{flushright}
CU-PHYSICS/1-2010
\end{flushright}
\begin{center}
{\Large {\bf
{Probing CPT Violation in B Systems}}}\\[5mm]
\bigskip
{\sf Anirban Kundu} $^a$,
{\sf Soumitra Nandi} $^b$, and
{\sf Sunando Kumar Patra} $^{a}$

\bigskip\bigskip

$^a${\footnotesize\rm
Department of Physics, University of Calcutta, \\
92, Acharya Prafulla Chandra Road, Kolkata 700 009, India. \\
E-mail: {\sf akphy@caluniv.ac.in, sunandoraja@gmail.com} }

\bigskip
$^b${\footnotesize\rm
Dip.\ Fisica Teorica, Univ.\ di Torino \& INFN Torino, I-10125, Torino, Italy\\
E-mail: {\sf nandi@to.infn.it} }

\normalsize
\vskip 10pt

{\large\bf Abstract}
\end{center}

\begin{quotation} \noindent

We discuss how a possible violation of the combined symmetry CPT in the
B meson system can be investigated at the LHC. We show
how a tagged and an untagged analysis of the decay modes of both $B_d$ and
$B_s$ mesons can
lead not only to a possible detection of a CPT-violating new physics but
also to an understanding of its precise nature. The implication of CPT
violation to a large mixing phase in the $B_s$ system is also discussed.
\vskip 10pt
PACS numbers: {\tt 11.30.Er, 14.40.Nd}\\
\end{quotation}
\begin{flushleft}\today\end{flushleft}
\vfill
\end{titlepage}
\newpage
\setcounter{page}{1}
\section{Introduction}
\label{sec1}

The combined symmetry CPT is supposed to be an exact symmetry of any local
axiomatic quantum field theory. This is indeed supported by the experiments:
all possible tests so far \cite{pdg} have yielded negative results, consistent
with no CPT violation. Why then should we be interested in the possibility of
CPT violation in the B system? There are three main reasons: first, any
symmetry which is supposed to be exact ought to be questioned and investigated,
and we may get a surprise, just like the discovery of CP violation; second,
it is not obvious that CPT will still be an exact symmetry in the bound
state of quarks and antiquarks, where the asymptotic states are not
uniquely defined \cite{quark}; third, there may be some nonlocal
and nonrenormalisable string-theoretic effects at the Planck scale which have a ramification at
the weak scale through the effective Hamiltonian \cite{cpt-hamil}.
Moreover, this effect can
very well be flavour-sensitive, and so the constraints obtained from the
K system \cite{nussinov} may not be applicable to the B systems. A comprehensive
study of CPT violation in the neutral K meson system, with a formulation
that is closely analogous to that in the B system, may be found in \cite{lavoura}.

There are already some investigations on CPT violation in B systems. Datta
{\em et al.} \cite{datta} have shown how CPT violation can lead to a significant
lifetime difference $\Delta\Gamma/\Gamma$ in the generic $M^0$-$\overline{M}
{}^0$ system, where $M^0=K^0, B^0$, or $B_s$. It was discussed in \cite{balaji}
how direct CP asymmetries and semileptonic decays can act as a
probe of CPT violation. Signatures of CPT violation in non-CP eigenstate channels
was discussed in \cite{xing1}. The role of dilepton asymmetry as a test of
CPT violation and possible discrimination from $\Delta B = - \Delta Q$ processes
were investigated in \cite{xing2}. The BaBar experiment at SLAC has tried to look for
CPT violation in the diurnal variations of CP-violating observables and set
some limits \cite{cpt-babar}.

Right now, there is no signature of CPT violation, or for that matter any
type of new physics, in the width difference of $\bbbar$ and decay channels of
$B_d$ \footnote{We use $B^0$ and $\overline{B}{}^0$ to indicate the
flavour eigenstates, $B_d$ as a generic symbol for both of them, and similarly
for $B_s$. The symbol $B_q$ will mean either a $B_d$ or a $B_s$.}.
The width difference for the $B_d$ system, $\Delta\Gamma_d$, is too small
yet to be measured experimentally, and the bound is
compatible with the Standard Model (SM). On the other hand, it is expected
that the width difference $\Delta\Gamma_s$ would be significant for the $B_s$
system, but at
the same time we know that the theoretical uncertainties are quite
significant \cite{lenz}. If there is some new physics (NP) that does not
contribute to the absorptive part of the $\bsbsbar$ box, the width difference
can only go down \cite{grossman}, while there are models where this conclusion
may not be true \cite{dighe}. To add to this murky situation, the CP-violating phase
$\beta_s$, which is expected to be very small from the CKM paradigm, has
been measured \cite{cdf-bs} to be large, compatible with the SM expectations
only at the $2.1\sigma$ level. The situation is interesting: there is hint of
some NP, but we are yet to be certain of its exact nature, not to mention
whether it exists at all.

In this situation, let us try to see what we can expect at the LHC, where
the $B_s$ meson, along with the $B_d$, will be copiously produced.
We are helped by the fact that the time resolution in ATLAS and
CMS are of the order of 40 fs, so one can track the time evolution of
even the rapidly oscillating
$B_s$. Thus, we expect excellent tagged and untagged measurements of both
$B_d$ and $B_s$ mesons. It is best to focus upon the single-amplitude
observables: $B_d\to J/\psi K_S$ and $B_s\to J/\psi\phi$ or $B_s\to D_s^+
D_s^-$ \footnote{They are not strictly single-channel as there is a penguin
process whose dominant part has the same phase as the leading Cabibbo-allowed
tree process, but on the other hand these channels are easy to measure, and
the penguin pollution is quite small and well under control.}.
For the $J/\psi\phi$ mode, one has to perform the angular
analysis and untangle the CP-even and CP-odd channels.

In this paper, we will discuss how one can detect the presence of a CPT
violating new physics from the tagged and untagged measurements of the
decay. We will confine our discussion to the case where CPT violation is
small compared to the SM amplitude, just to make the results more transparent.
The conclusions do not change qualitatively if the CPT violation is large,
which, we must say, is a far-off possibility based on the data from the
other experiments \cite{cpt-babar}.
We will also show how the nature of the CPT violating term
can be probed through these measurements.

In Section 2, we mention the relevant expressions, and introduce CPT violation,
with relevant expressions, in Section 3. The analysis for both $B_d$ and $B_s$
systems is performed in Section 4, while we summarise and conclude in Section
5.

\section{Basic Formalism}

Let us introduce CPT violation in the Hamiltonian matrix through the
parameter $\delta$ which can potentially be complex:
\be
\delta = \frac{H_{22}-H_{11}}{\sqrt{H_{12}H_{21}}}\,,
\ee
so that
\be
  \mathcal M = \left[\left(
 \begin{array}{cc}
 M_0-\delta' & M_{12}\\
 M_{12}^* & M_0+\delta'
 \end{array}
\right) - \frac{i}{2} \left(
 \begin{array}{cc}
 \Gamma_0 & \Gamma_{12}\\
 \Gamma_{12}^* & \Gamma_0
 \end{array}
\right)\right]\,,
\ee
where $\delta'$ is defined by
\be
\delta = \frac{2\delta'}{\sqrt{H_{12}H_{21}}}\,.
\ee

Solving the eigenvalue equation of $\mathcal M$, we get,

\bea
 \lambda &=& \left(M_0- \frac{i}{2}\Gamma_0\right) \pm H_{12}\alpha y
\nonumber
\\
{\rm or},\ \   \lambda &=& \left[H_{11} + H_{12}\alpha \left(y +
\frac{\delta}{2}\right)\right]\,,\ \
 \left[H_{22} - H_{12}\alpha \left(y + \frac{\delta}{2}\right)\right]\,,
\eea
where
$y = \sqrt{ 1 + \frac{\delta^2}{4} }$ and
$\alpha = \sqrt{ H_{21} / H_{12} }$.

Hence, corresponding eigenvectors or the mass eigenstates are defined as
\bea
 |B_H\ket & =& p_1|B^0\ket + q_1|\overline{B}{}^0\ket\,,
\nonumber\\
|B_L\ket &=& p_2|B^0\ket - q_2|\overline{B}{}^0\ket\,.
\eea
The normalisation satisfies
\begin{equation}
|p_1|^2 + |q_1|^2 = |p_2|^2 + |q_2|^2 = 1\,.
\end{equation}

Let us define,
\be
 \eta_1 = \frac{q_1}{p_1} = \left(y + \frac{\delta}{2}\right) \alpha\,;
\ \ \
\eta_2 = \frac{q_2}{p_2} = \left(y - \frac{\delta}{2}\right) \alpha\,;
\ \ \
\omega = \frac{\eta_1}{\eta_2}\,.
\label{fac1}
\ee
The convention of \cite{cpt-babar} leads to $z_0=\delta/2$, where $z_0$ is a measure of CPT violation as
used in \cite{cpt-babar}. The limits imply that $|z_0|\ll 1$. Even if the origin of CPT violation is something
different, it is not unrealistic to assume $|\delta| \ll 1$.

One could even relax the assumption of $H_{21}=H_{12}^\ast$. However, there are two points that one
must note. First, the effect of expressing
$H_{12} = h_{12}+\bar\delta$, $H_{21} = h_{12}^\ast - \bar\delta$ appears as ${\bar\delta}^2$ in
$\sqrt{H_{12}H_{21}}$, the relevant expression in eq.\ (1), and can be neglected if we assume
$\bar\delta$ to be small. The second point, which is more important, is that CPT conservation constrains only
the diagonal elements and puts no constraint whatsoever on the off-diagonal elements.
It has been shown in \cite{lavoura,balaji} that $H_{12}\not= H_{21}^\ast$ leads to
T violation, and only $H_{11}\not= H_{22}$ leads to unambiguous CPT violation. Thus, we will
focus on the parametrization used in eqs.\ (1) and (2) to discuss the effects of CPT violation.

The time-dependent flavour eigenstates are given by
\bea
|B_q(t)\ket & = &f_+(t) |B_q\ket + \eta_1 f_-(t) |\overline{B_q}\ket \nonumber\\
|\overline{B_q}(t)\ket & =& \frac{f_-(t)}{\eta_2} |B_q\ket + \bar f_{+}(t)
|\overline{B_q}\ket\,,
\eea
where
\bea
\nonumber f_-(t) & = &
\frac{1}{(1 + \omega)} \left(e^{-i\lambda_1 t} - e^{-i\lambda_2 t}\right)\,, \\
\nonumber f_+(t) & = &\frac{1}{(1 + \omega)} \left(e^{-i\lambda_1 t} + \omega e^{-i\lambda_2 t}\right)\,, \\
 \bar f_+(t) & = &\frac{1}{(1 + \omega)} \left(\omega e^{-i\lambda_1 t} + e^{-i\lambda_2 t}\right)\,.
\label{funct}
\eea

So, the decay rate of the meson $B_q$ at time $t$ to a CP eigenstate $f$
is given by
\bea
\Gamma(B_q(t)\rightarrow f_{CP}) & = &
\left[|f_+(t)|^2 + |\xi_{f_1}|^2 |f_-(t)|^2 + 2{\rm Re}\left(\xi_{f_1} f_-(t) f_+^*(t)\right)\right] |A_f|^2\,, \nonumber\\
\Gamma(\overline{B_q}(t)\rightarrow f_{CP}) & = &
\left[|f_-(t)|^2 + |\xi_{f_2}|^2 |\bar f_+(t)|^2 + 2{\rm Re}\left(\xi_{f_2} \bar f_+(t) f_-^*(t)\right)\right] \left|\frac{A_f}{\eta_2}\right|^2\,,
\label{decay1}
\eea
where
\be
 A_f = \bra f|H|B_q\ket\,,\ \ \
\bar A_f = \bra f|H|\overline{B_q}\ket\,.
\ee
Also,
\be
\xi_{f_1} = \eta_1 \frac{\bar A_f}{A_f}\,,\ \ \
\xi_{f_2} = \eta_2 \frac{\bar A_f}{A_f}\,.
\ee
In the SM, both are equal and $\xi_{f_1}=\xi_{f_2}=\xi_f$. For single-channel processes, $|\xi_f|=1$.

%
%
%

Now using eq.\ (\ref{fac1}) and eq.\ (\ref{funct}) one gets
\bea
 \left|f_-(t)\right|^2 & = &
\frac{2 e^{-\Gamma t}}{|1+\omega|^2} \left[\cosh\left(\frac{\Delta\Gamma t}{2}
\right) - \cos\left(\Delta{m} t\right)\right]\,,\nonumber\\
 \left|f_+(t)\right|^2 & = &
\frac{e^{-\Gamma t}}{|1+\omega|^2} \Bigg[ \cosh\left(\frac{\Delta \Gamma t}{2}
\right)(1+|\omega|^2) + \sinh\left(\frac{\Delta \Gamma t}{2}
\right)(1-|\omega|^2)  \nonumber \\
&& + 2 {\rm Re}(\omega) \cos\left(\Delta m t\right) - 2 {\rm Im}(\omega)
\sin\left(\Delta m t\right)\Bigg]\,, \nonumber \\
 \left|\bar f_+(t)\right|^2 & =&
  \frac{e^{-\Gamma t}}{|1+\omega|^2} \Bigg[ \cosh\left(\frac{\Delta \Gamma t}{2}
\right)(1+|\omega|^2) - \sinh\left(\frac{\Delta \Gamma t
}{2}\right)(1-|\omega|^2)  \nonumber \\
&& + 2{\rm Re}(\omega) \cos\left(\Delta m t\right) + 2 {\rm Im}(\omega)
\sin\left(\Delta m t\right)\Bigg]\,,\nonumber \\
 f_+^*(t) f_- (t) & =& \frac{e^{-\Gamma t}}{|1+\omega|^2}
\Bigg[ \cosh\left(\frac{\Delta \Gamma t}{2}\right)(1-\omega^{\ast}) +
\sinh\left(\frac{\Delta \Gamma t }{2}\right)(1+\omega^{\ast})  \nonumber \\
&& + \cos\left(\Delta m t\right) (-1 + \omega^{\ast}) - i \sin
\left(\Delta m t\right)(1+ \omega^{\ast}) \Bigg]\,, \nonumber \\
\bar{f_+}(t) f^{\ast}_{-} (t) & =&
\frac{e^{-\Gamma t}}{|1+\omega|^2} \Bigg[ \cosh\left(\frac{\Delta \Gamma t}{2}
\right)(\omega-1) +
\sinh\left(\frac{\Delta \Gamma t }{2}\right)(1+\omega)  \nonumber \\
&& + \cos\left(\Delta m t\right) (1 - \omega) +
i \sin\left(\Delta m t\right)(1+ \omega) \Bigg]\,.
\label{fac2}
\eea
Here, $\Delta m$ and $\Delta \Gamma$ are defined through;
\begin{equation}
\lambda_1-\lambda_2 = \Delta m + \frac{i}{2}\Delta \Gamma\,,
\end{equation}
with
\be
\lambda_{(1,2)} = m_{(1,2)} - \frac{i}{2} \Gamma_{(1,2)}\,, \ \ \
\Delta{m} = m_1-m_2\,, \ \ \
\Delta\Gamma = \Gamma_2 - \Gamma_1\,.
\ee

\section{Introducing CPT Violation}

If we consider a time-independent CPT violation so that $\delta$ is a constant, there are only
two unknowns in the picture: Re($\delta$) and Im($\delta$), over those in the SM. We will try to see
how one can extract informations about them. For our analysis, let us take $\delta$ to be
complex; it will be straightforward to go to the simpler limiting cases where $\delta$ is purely
real or imaginary.  For example, if the width difference $\Delta\Gamma$ is much smaller
than $\Delta m$, the model of \cite{cpt-babar} makes $\delta$ completely real.

When $B_q$ and $\bar{B}_q$ are produced in equal numbers, the untagged decay rate can be defined as
\begin{align}
\Gamma_U[f,t]=\Gamma(B_q(t)\to f) + \Gamma(\bar{B}_q(t)\to f)\,,
\label{untag}
\end{align}
using the above expression one could define the branching fraction as
\begin{align}
Br[f]= \frac{1}{2}\int^{\infty}_{0}{dt ~\Gamma[f,t]}\,.
\label{branch}
\end{align}
The above equation is useful to fix the overall normalization.

We assume, $\delta \ll 1$ and expand any function $f(\delta)$ using Taylor series expansion and drop all the terms ${\cal{O}}(\delta^n)$ for $n >2$.
From eq. (\ref{untag}), eq. (\ref{decay1}) and eq. (\ref{fac2}) we will get the untagged decay rate for the decay $B_q\to f$,
\begin{align}
\Gamma_U[f,t] &= |A_f|^2 e^{-\Gamma_q t} \Bigg[ \left\{(1+ |\xi_f|^2)(1+ \frac{({\rm Im}(\delta))^2}{4})- {\rm Im}(\delta) {\rm Im}(\xi_f)\right\} \cosh\left(\frac{\Delta\Gamma_q t}{2}\right) \nonumber \\
& \hskip 60pt - \left\{(1+ |\xi_f|^2)\frac{({\rm Im}(\delta))^2}{4}- {\rm Im}(\delta) {\rm Im}(\xi_f)\right\} \cos\left(\Delta m_q t\right) \nonumber \\
& \hskip 60pt + \left\{2 {\rm Re}(\xi_f)- \frac{1}{2}(1- |\xi_f|^2){\rm Re}(\delta)-\frac{1}{4} {\rm Re}(\xi_f)(({\rm Re}(\delta))^2- ({\rm Im}(\delta))^2)\right\} \times \nonumber \\
& \hskip 60pt \sinh\left(\frac{\Delta\Gamma_q t}{2}\right)- \frac{1}{2}{\rm Im}(\delta)\left\{(1- |\xi_f|^2)+ {\rm Re}(\delta){\rm Re}(\xi_f)\right\} \sin\left(\Delta m_q t\right)\Bigg]\,.
\end{align}

Thus, for the $B_s$ system, where the hyperbolic functions are not negligible, we get (keeping up to first order of
terms in $\Delta\Gamma_s$):
\begin{align}
Br[f] &= \frac{1}{2}\int^{\infty}_{0}{dt~ \Gamma[f,t]}\nonumber\\
&= \frac{|A_f|^2}{2} \Bigg[ \frac{1}{\Gamma_s}\left\{(1+ |\xi_f|^2)(1+ \frac{({\rm Im}(\delta))^2}{4})- {\rm Im}(\delta) {\rm Im}(\xi_f)\right\}  \nonumber \\
& \hskip 40pt - \frac{\Gamma_s}{(\Delta m)^2+ (\Gamma_s)^2}\left\{(1+ |\xi_f|^2)\frac{({\rm Im}(\delta))^2}{4}- {\rm Im}(\delta) {\rm Im}(\xi_f)\right\} \nonumber \\
& \hskip 40pt + \frac{\Delta \Gamma_s}{2 (\Gamma_s)^2}\left\{2 {\rm Re}(\xi_f)- \frac{1}{2}(1- |\xi_f|^2){\rm Re}(\delta)-\frac{1}{4} {\rm Re}(\xi_f)(({\rm Re}(\delta))^2- ({\rm Im}(\delta))^2)\right\}  \nonumber \\
& \hskip 40pt - \frac{1}{2}{\rm Im}(\delta)\left\{(1- |\xi_f|^2)+ {\rm Re}(\delta){\rm Re}(\xi_f)\right\} \frac{\Delta m}{(\Delta m)^2+ (\Gamma_s)^2}\Bigg]
\label{br-eq1}
\end{align}

Theoretically, one can obtain the coefficients of the trigonometric and the hyperbolic terms by fitting
the untagged decay rate. In actual cases this is a difficult task. However, there is one other observable which
may help us. Before we go to that, let us note that the
above expression is further simplified in the following four cases.
\begin{itemize}
\item For the $B_d$ system: We can neglect $\Delta\Gamma_d$
so that the cosh term is unity and the sinh term is zero. Thus, there are only two time-dependent
terms, $\cos(\Delta mt)$ and $\sin(\Delta mt)$, and the fitting is easier. Note that $\Delta\Gamma_d$ is
measured to be small, so we need not consider the case where it is enhanced to a significant value
because of the CPT violation. In fact, if $\delta$ is small, $\Delta\Gamma_d$ is bound to be that
coming from the SM, as the correction is proportional only to $\delta^2$ and higher.

\item For one-amplitude processes: We can put
$|\xi_f|=1$, and only one of Re($\xi_f$) and Im($\xi_f$) remains a free parameter
\footnote{$\xi_f$ is a SM quantity, so it is theoretically calculable, but it may also contain other new physics
which is CPT conserving, so it is better to obtain both real and imaginary parts of $\xi_f$ and
check whether $|\xi_f|=1$.}.

\item For $\delta$ being either purely real or purely imaginary: The expressions are straightforward.
For example, if $\delta$ is purely real, there is no trigonometric dependence on the untagged
rate.

\item Finally, for $|\delta| \ll 1$:  We can neglect terms higher than linear in either Re($\delta$) or Im($\delta$)
in eq.\ (\ref{br-eq1}).
This is expected to be the case according to \cite{cpt-babar}. For example, the expression for the branching
fraction for a one-amplitude process simplifies to
\be
Br[f]
= \frac{|A_f|^2}{2} \Bigg[ \frac{1}{\Gamma_s}\left\{2- {\rm Im}(\delta) {\rm Im}(\xi_f)\right\}  +
 \frac{\Gamma_s}{(\Delta m)^2+ (\Gamma_s)^2} {\rm Im}(\delta) {\rm Im}(\xi_f)+ \frac{\Delta \Gamma_s}{(\Gamma_s)^2} {\rm Re}(\xi_f)
\Bigg]\,.
\label{br-eq2}
\ee

\end{itemize}

One can also tag the B mesons and define a tagged decay rate asymmetry
\begin{align}
\Gamma_T[f,t] &= \Gamma(B_q(t)\to f) - \Gamma(\bar{B}_q(t)\to f) \nonumber \\
& = |A_f|^2 e^{-\Gamma_q t} \Bigg[ \left\{(1- |\xi_f|^2)\frac{({\rm Re}(\delta))^2}{4}- {\rm Re}(\delta) {\rm Re}(\xi_f)\right\} \cosh\left(\frac{\Delta\Gamma_q t}{2}\right) \nonumber \\
& \hskip 60pt + \left\{(1 - |\xi_f|^2)(1- \frac{({\rm Re}(\delta))^2}{4}) + {\rm Re}(\delta){\rm Re}(\xi_f)\right\}\cos\left(\Delta m_q t\right) \nonumber \\
& \hskip 60pt - \frac{1}{2}{\rm Re}(\delta)\left\{(1+|\xi_f|^2)- {\rm Im}(\delta){\rm Im}(\xi_f)\right\}\sinh\left(\frac{\Delta\Gamma_q t}{2}\right)+ \Big\{2 {\rm Im}(\xi_f) \nonumber \\
& \hskip 60pt - \frac{1}{2}{\rm Im}(\delta)(1 +|\xi_f|^2)- \frac{1}{4} {\rm Im}(\xi_f)(({\rm Re}(\delta))^2- ({\rm Im}(\delta))^2)\Big\} \sin\left(\Delta m_q t\right)\Bigg]\,.
\label{untag5}
\end{align}

Note that (i) for Re($\delta$)=Im($\delta$)=0, this reverts back to the SM expression, as it should, and
(ii) If $|\delta|\ll 1$ and $\Delta\Gamma/\Gamma \ll 1$ as in the $B_d$ system, the tagged rate can measure
both Re($\delta$) and Im($\delta$).

For one-amplitude processes with $|\delta|\ll 1$, one may write a simplified expression:
\begin{align}
\Gamma_U[f,t] &=  |A_f|^2 e^{-\Gamma_q t}\Bigg[ (2 - {\rm Im}(\delta) {\rm Im}(\xi_f)) \cosh\left(\frac{\Delta\Gamma_q t}{2}\right) \nonumber \\
& \hskip 60pt + {\rm Im}(\delta){\rm Im}(\xi_f)\cos\left(\Delta m_q t\right)
+ 2 {\rm Re}(\xi_f) \sinh\left(\frac{\Delta\Gamma_q t}{2}\right)
\Bigg]\,, \nonumber \\
\Gamma_T[f,t] & =|A_f|^2 e^{-\Gamma_q t}\Bigg[ - {\rm Re}(\delta) {\rm Re}(\xi_f) \cosh\left(\frac{\Delta\Gamma_q t}{2}\right) + {\rm Re}(\delta){\rm Re}(\xi_f)\cos\left(\Delta m_q t\right) \nonumber \\
& \hskip 60pt - {\rm Re}(\delta) \sinh\left(\frac{\Delta\Gamma_q t}{2}\right)  + \left\{2 {\rm Im}(\xi_f) - {\rm Im}(\delta) \right\} \sin\left(\Delta m_q t\right)\Bigg]\,.
\label{untag6}
\end{align}
It is clear from eq.\ (\ref{untag6}) how one can extract ${\rm Re}(\delta)$ and
${\rm Im}(\delta)$ by comparing the untagged and tagged analyses. Suppose weconsider the $B_s$ system where $\Delta\Gamma_s$ is non-negligible. The coefficient
of the sinh term in $\Gamma_T$ gives ${\rm Re}(\delta)$. However, there is an
overall normalisation uncertainty given by $|A_f|^2$. To remove this, one can consider
 a combined study of the coefficients of $\sinh\left(\frac{\Delta\Gamma_s t}{2}\right)$ and
$\cos\left(\Delta m_s t\right)$ from the untagged and tagged decay rates respectively; their
ratio allows for a clean extraction of ${\rm Re}(\delta)$.  On the other
hand, the ratio of the coefficients of
$\cos(\Delta m_st)$ in $\Gamma_U$ and $\sin(\Delta m_st)$ in $\Gamma_T$
gives a clean measurement of ${\rm Im}(\delta)$, as ${\rm Im}(\xi_f)$ is known
from the SM dynamics. A further check about the one-amplitude nature is provided
from $|{\rm Re}(\xi_f)|^2 + |{\rm Im}(\xi_f)|^2 = 1$. In fact, as long as $\delta$ is small,
one can extract both ${\rm Re}(\delta)$ and ${\rm Im}(\delta)$ even if $|\xi_f|\not= 1$,
from the coefficients of the sine, cosine, and hyperbolic sine terms in $\Gamma_U$
and $\Gamma_T$.

One may also define the time-dependent CPT asymmetry as
\be
A_{CPT}(f,t) = \frac{\Gamma_T[f,t]}{\Gamma_U[f,t]}
= \frac{ \Gamma(B_q(t)\to f) - \Gamma(\bar{B}_q(t)\to f) }
{ \Gamma(B_q(t)\to f) + \Gamma(\bar{B}_q(t)\to f) }\,,
\ee
and the time-independent CPT asymmetry as
\be
A_{CPT}(f) = \frac{\int_0^\infty dt~\Gamma_T[f,t]}{\int_0^\infty dt~\Gamma_U[f,t]}
= \frac{ \int_0^\infty dt~[\Gamma(B_q(t)\to f) - \Gamma(\bar{B}_q(t)\to f)] }
{ \int_0^\infty dt~[\Gamma(B_q(t)\to f) + \Gamma(\bar{B}_q(t)\to f)] }
\,.
\ee
This goes to the usual CP asymmetry $A_{CP}$ if $\delta=0$; thus, any deviation from the expected CP
asymmetry calculated from the SM would signal new physics, but one must check all the boxes to
pinpoint the nature of the new physics. For example, there would not be any change in the semileptonic
CP asymmetry if the new physics is only CPT violating in nature.

\section{Analysis}

There are five {\em a priori} unknowns: ${\rm Re}(\delta)$, ${\rm Im}(\delta)$, ${\rm Re}(\xi_f)$, ${\rm Im}(\xi_f)$,
and $|A_f|^2$. For a one-amplitude process $|\xi_f|^2=1$ and the number of unknowns reduce to four. The
tagged and untagged decay rates, the branching fraction, and the time-independent CPT asymmetry would
provide informations on all of these unknowns. Assuming the CPT-conserving physics to be purely that
of the SM, one may calculate $\xi_f$ following the CKM picture. In the analysis that follows, we take $\xi_f$
to be known from the SM. We would like to point out the following features.

\begin{itemize}
\item
The overall amplitude $|A_f|^2$ cancels in the CPT asymmetry. This, therefore, is going to be the observable one
needs to measure most precisely.

\item
It is enough to measure the coefficients of the trigonometric terms only. For the $B_d$ system, $\Delta\Gamma_d$
is small anyway, and for the $B_s$ system, $\Delta\Gamma_s$ has a large theoretical uncertainty.

\item
The analysis holds even if the process under consideration is not a one-amplitude process. In fact, one may
check whether there is a second CPT conserving new physics amplitude just by looking at the extracted
values of ${\rm Re}(\xi_f)$ and ${\rm Im}(\xi_f)$.

\item
The coefficient of $\sin(\Delta m_qt)$ in the expression for the tagged decay rate $\Gamma_T$ gives the
mixing phase in the box diagram. Thus, ${\rm Im}(\delta)$ may be constrained by the CP asymmetry
measurements in the $B_d$ system. On the other hand, even those constrained values generate a large
mixing phase for the $B_s$ system compatible with the CDF data.
\end{itemize}

\subsection{The $B_s$ system}

\begin{figure}[htbp]
\vspace{-10pt}
\centerline{\hspace{-3.3mm}
\rotatebox{0}{\epsfxsize=12cm\epsfbox{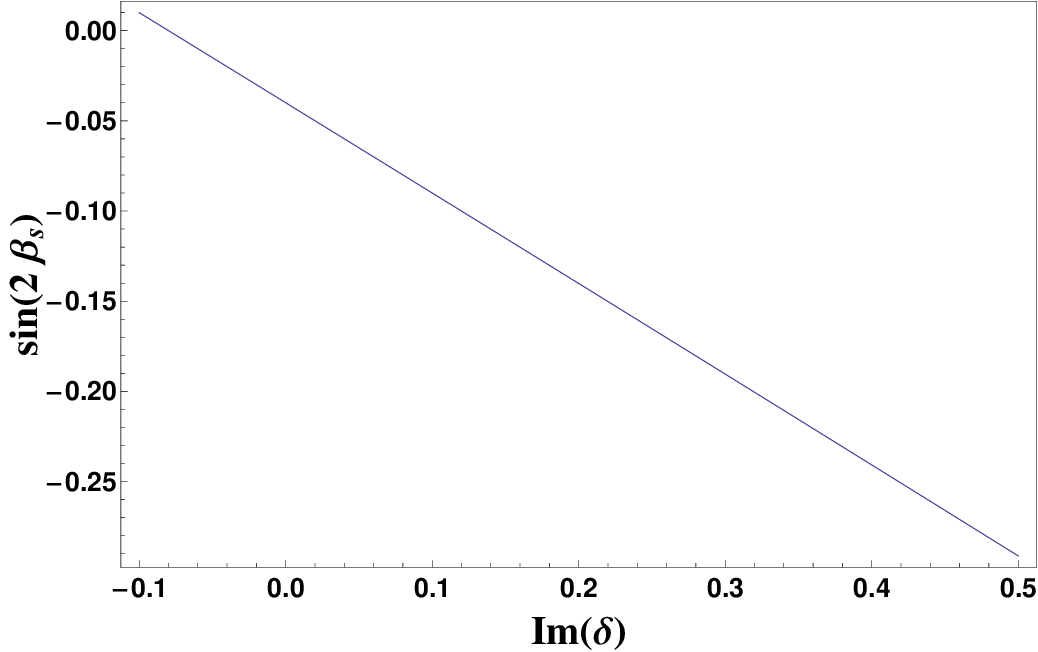}}}
\caption{Variation of $\sin(2\beta_s)$ with ${\rm Im}(\delta)$.}
\label{s2betas-imd}
\end{figure}

For the $B_s$ system, we take
\bea
&&\Delta m_s = 17.77\pm 0.12 {\rm ps}^{-1}\,,\
 \Delta \Gamma_s =0.096\pm 0.039 {\rm ps}^{-1}\,, \
 \frac{\Delta \Gamma_s}{\Gamma_s} = 0.147\pm 0.060\,, \nonumber\\
 &&\frac{1}{\Gamma_s} = 1.530\pm 0.009 {\rm ps}\,, \
 {\rm Re}(\xi_f)=0.99\,,\
 {\rm Im}(\xi_f) = -0.04\,.
 \eea
\begin{figure}[htbp]
\vspace{-10pt}
\centerline{\hspace{-3.3mm}
\rotatebox{0}{\epsfxsize=12cm \epsfbox{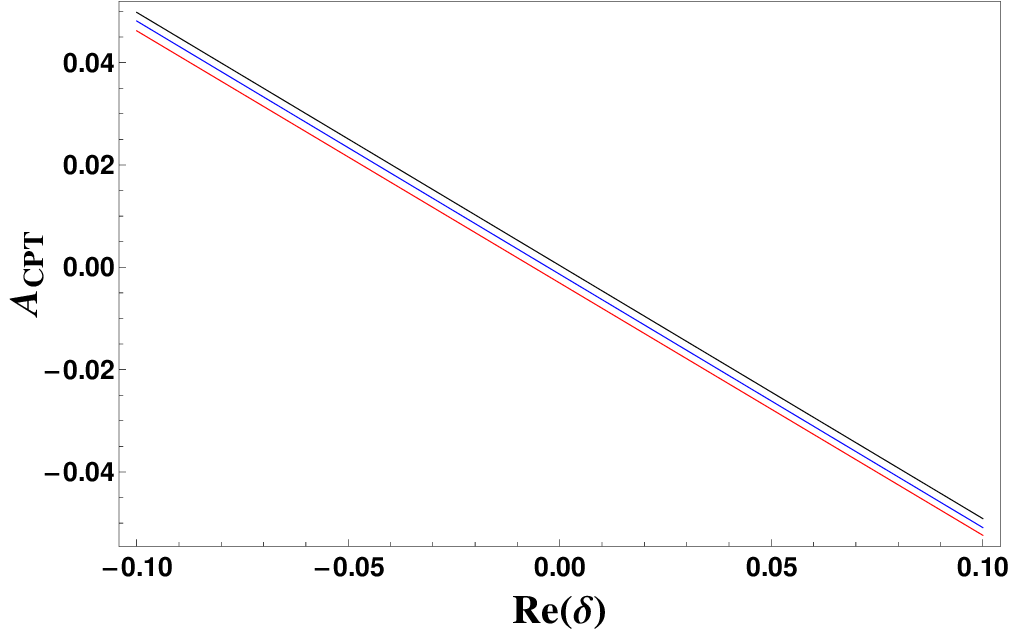}}}
\caption{Variation of $A_{CPT}$ with ${\rm Re}(\delta)$ for the $B_s$ system.
The three lines, from top to bottom,
are for ${\rm Im}(\delta) = -0.1, 0$ and $0.1$ respectively.}
\label{acpt-bs}
\end{figure}
In figure \ref{acpt-bs}, we show the variation of $A_{CPT}$ with ${\rm Re}(\delta)$. For our analysis, we take
both $|{\rm Re}(\delta)|, |{\rm Im}(\delta)| < 0.1$, which is consistent with \cite{cpt-babar}. The variation of
$A_{CPT}$ with $\Delta m_s$ and $\Delta\Gamma_s$ is negligible, of the order of 0.2\%, so we fix
them to their respective central values. Effects of $\delta$ in both $\Delta m_s$ and $\Delta\Gamma_s$
are quadratic in $\delta$, and hence we can use the SM values for them. In fact, $A_{CPT}$ does not
depend significantly on the choice of ${\rm Im}(\delta)$ either; the variation is less than 1\%.
This is due to the fact that here, $|{\rm Im}(\xi_f)| \ll |{\rm Re}(\xi_f)|$ and hence the coefficient of
${\rm Re}(\delta)$ is much greater than the coefficient of ${\rm Im}(\delta)$ in the expression of
$A_{CPT}$. This feature does not hold for the $B_d$ system. Note that
$A_{CPT}$ clearly gives the sign of ${\rm Re}(\delta)$.
The small nonzero value of $A_{CPT}$
for $\delta=0$ indicates the small mixing phase in the $\bsbsbar$ box diagram.
However, the apparent phase, i.e., the coefficient of $\sin(\Delta m_st)$, can increase with ${\rm Im}(\delta)$,
as can be seen from figure \ref{s2betas-imd}.

\subsection{The $B_d$ system}
The inputs that we use for the $B_d$ system are
\be
\Delta m_d = 0.507 {\rm ps}^{-1}\,,\
 \Delta \Gamma_d =0\,,\
 {\rm Re}(\xi_f)=0.72\,,\
 {\rm Im}(\xi_f) = 0.695\,.
 \ee
\begin{figure}[htbp]
\vspace{-10pt}
\centerline{\hspace{-3.3mm}
\rotatebox{0}{\epsfxsize=12cm\epsfbox{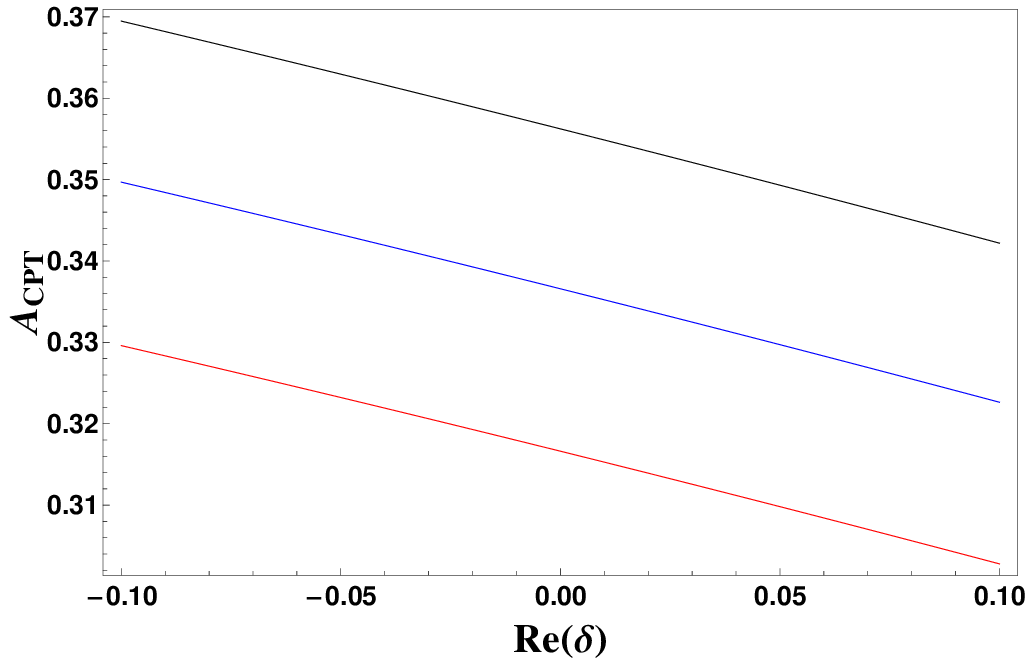}}}
\hspace{3.3cm}
\caption{Variation of $A_{CPT}$ with ${\rm Re}(\delta)$ for the $B_d$ system.
The three lines, from top to bottom,
are for ${\rm Im}(\delta) = -0.1, 0$ and $0.1$ respectively.}
\label{acpt-bd1}
\end{figure}


\begin{figure}[htbp]
\vspace{-10pt}
\centerline{\hspace{-3.3mm}
\rotatebox{0}{\epsfxsize=12cm\epsfbox{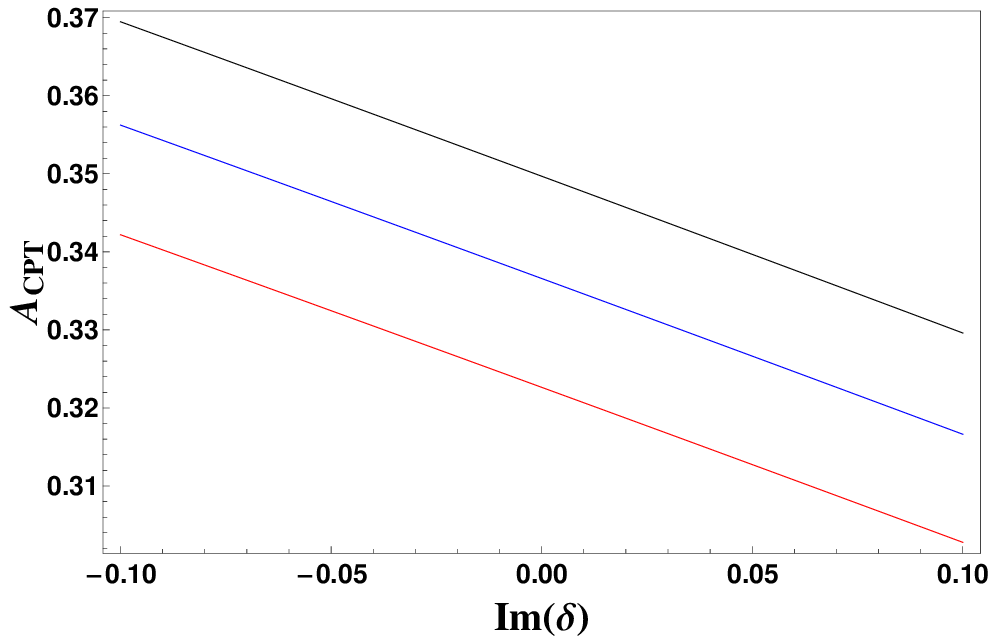}}}
\hspace{3.3cm}
\caption{Variation of $A_{CPT}$ with ${\rm Im}(\delta)$ for the $B_d$ system.
The three lines, from top to bottom,
are for ${\rm Re}(\delta) = -0.1, 0$ and $0.1$ respectively.}
\label{acpt-bd2}
\end{figure}
\begin{figure}[htbp]
\vspace{-10pt}
\centerline{\hspace{-3.3mm}
\rotatebox{0}{\epsfxsize=12cm\epsfbox{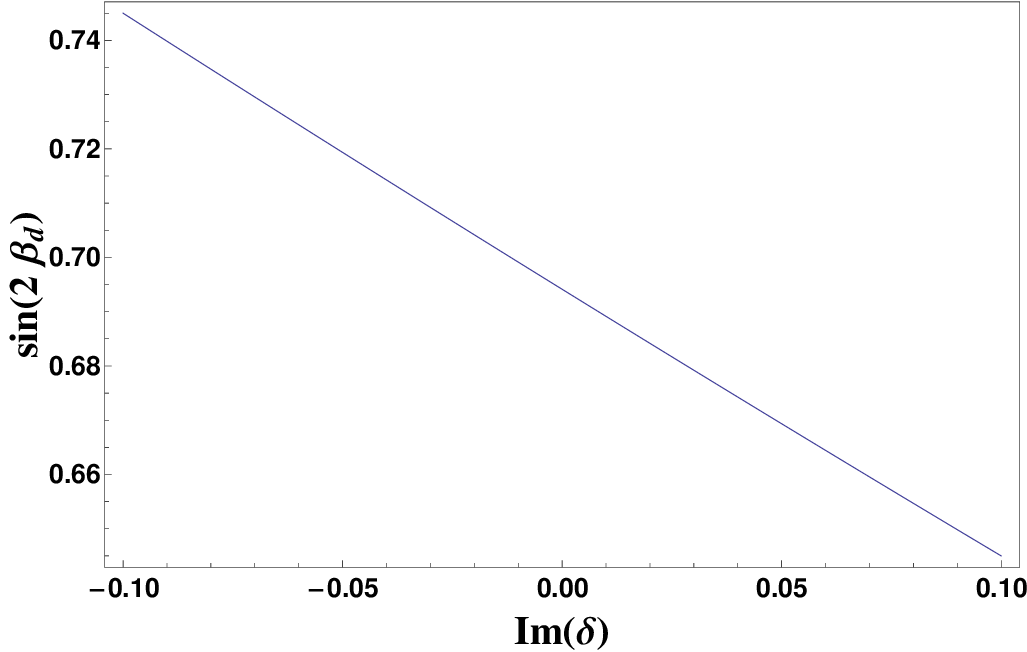}}}
\hspace{3.3cm}
\caption{Variation of $\sin(2\beta_d)$ with ${\rm Im}(\delta)$.}
\label{s2betad-imd}
\end{figure}

This follows from the CKM expectation of $\sin(2\beta_d)=0.695\pm 0.020$. The constraint on $\delta$ comes
from the measurement of $\sin(2\beta_d)$ in the $b\to c\bar{c}s$ channel: $0.668 \pm 0.028$ \cite{utfit}
\footnote{We do not take the measurements coming from $b\to s$ penguin channels because of their
inherent uncertainties.}.
Again, we can fix $\Delta m_d$ at its central value. This time, due to the comparable values of
Re($\xi_f$) and Im($\xi_f$), $A_{CPT}$ is sensitive to both Re($\delta$) and Im($\delta$). The variations
are shown in figure \ref{acpt-bd1} for three values of Im($\delta$) and figure \ref{acpt-bd2} for three
values of Re($\delta$). It turns out that $A_{CPT}$ is always positive for ${\rm Re}(\delta), {\rm Im}(\delta)
< 1$; this is a consistency check for the CPT violation.
Note that the measured value of $\sin(2\beta_d)$ can go down from its CKM
expectation for ${\rm Im}(\delta)>0$, in fact, for ${\rm Im}(\delta)\approx 0.07$, $\sin(2\beta_d)\approx
0.66$, as can be seen from figure \ref{s2betad-imd}. While this value of ${\rm Im}(\delta)$ generates a
mixing phase for the $B_s$ system that is consistent with the CDF and D0 measurements at $1\sigma$,
one must remember that $\delta$ need not be a flavour-blind parameter.

\section{Summary and Conclusions}

We have investigated the possibility of CPT violation in neutral B systems. CPT is a symmetry that is
expected to be exact and the violation, even if it exists, should be quite small. However, it is possible
to measure even a small CPT violation from the tagged and untagged decay rates of the neutral B
mesons. In particular, for single-amplitude decay channels, the coefficients of the trigonometric terms
$\sin(\Delta mt)$ and $\cos(\Delta mt)$ can effectively pinpoint the nature of the CPT violating
parameter $\delta$. This is an interesting possibility for the decays $B_s\to D_s^+ D_s^-$ and $B_S
\to J/\psi \phi$ (with an angular analysis). Even a small CPT violation, allowed by the mixing
constraints for the $B_d$ system, can make the $B_s$ mixing phase more compatible with the
Tevatron measurements, at the level of about $1\sigma$.
On the other hand
CPT violation should not affect the semileptonic CP asymmetries, as the corrections are
quadratic in nature, and expected to be negligible for small $\delta$. Thus, a correlated study of
the CP asymmetries in $B_s\to J\psi\phi$ and  $B_s\to D_s^+ D_s^-$ vis-a-vis
$B_s\to D_s \ell\nu$ might be useful to pinpoint the  CPT violating effects.
This, we feel, is something that the
experimentalists should look for in the coming years at the LHC.

\centerline{\bf{Acknowledgements}}

SKP acknowledges CSIR, Govt.\ of India, for a research fellowship.
SN would like to thank Ulrich Nierste for useful discussions. His work is supported by a European Community's Marie-Curie Research Training Network under contract MRTN-CT-2006-035505 ``Tools and Precision Calculations for Physics Discoveries at Colliders". The work of AK was supported by BRNS, Govt.\ of
India; CSIR, Govt.\ of India; and the DRS programme of the University Grants Commission.

\end{document}